\documentclass[12pt]{article}
\def\la{\mathrel{\mathpalette\fun <}}

\def\fun#1#2{\lower3.6pt\vbox{\baselineskip0pt\lineskip.9pt
\ialign{$\mathsurround=0pt#1\hfil##\hfil$\crcr#2\crcr\sim\crcr}}}

\title{$B_d(\bar B_d) \to \rho^\pm \pi^\mp, \rho^+ \rho^-, \pi^+ \pi^-$:
hunting for alpha}
\author{M.I. Vysotsky \\ ITEP, Moscow, Russia}
\date{}

\begin{document}
\maketitle

\begin{abstract}

We determine the domains of the values of unitarity triangle angle
$\alpha$, allowed by the charmless strangeless $B_d (\bar B_d)$
decays.
\end{abstract}

\section{Introduction}

In paper \cite{1} from the data on CP asymmetries in $B_d (\bar
B_d) \to \rho^\pm \pi^\mp$, $\rho^+ \rho^-$ decays and BABAR data
on CP asymmetries in $B_d(\bar B_d) \to \pi^+ \pi^-$ decays we
determine the value of angle $\alpha$ of the unitarity triangle:
\begin{equation}
\alpha = 96^o \pm 3^o \;\; , \label{1}
\end{equation}
where only a tree quark decay amplitude $\bar b \to u \bar u \bar
d (b \to u\bar u d)$ was taken into account. The numerical values
of angle $\alpha$ obtained from the considered decays are
consistent with each other and with the value which follows from
the global CKM fit. This observation testifies to the validity of
a proposed approach.

As the next step in the present paper we will study  what changes
in the values of $\alpha$ are induced by QCD penguins. Our aim is
twofold. First, in this way we will get an estimate of the
theoretical uncertainty of the value of $\alpha$ determined in
paper \cite{1}. Secondly, we will get the formulas for CP
violating parameters describing these decays which vanish when
penguins are neglected ($C_{\rho\pi}$, $A_{CP}^{\rho\pi}$,
$C_{\rho\rho}$, $C_{\pi\pi}$).

The angle shifts $\Delta\alpha$ we are interested in were
estimated in paper \cite{2}; however, in that paper FSI phases
were neglected, that is why $C_{\rho\pi} = A_{CP}^{\rho\pi} =
C_{\rho\rho} = C_{\pi\pi} = 0$ follows from \cite{2} (the nonzero
penguin amplitudes are a necessary, but not a sufficient condition
for $C_{\rho\pi} ...$ to be nonzero). We will take these phases
into account. The asymmetries depend on the differences of FSI
phases in the processes described by the tree and penguin
diagrams. One source of these differences is an imaginary part of
the quark penguin diagram, the so-called BSS mechanism of the
strong phases generation \cite{3} (see also paper \cite{4}). The
phase of the penguin diagram depends on the gluon $q^2$ which is
transferred to $u\bar u$ pair, each quark of which goes to
different $\pi^\pm$- or $\rho^\pm$-mesons. In this way the value
of $q^2$ depends on the light meson wave functions and we can
estimate it only roughly. Another source of FSI phases is hadron
rescattering and even less is known about the values of the phase
shifts between the penguin and tree diagrams generated in this
way. In view of this we will determine FSI phases from the
experimental data on CPV asymmetries, and investigate to what
values of $\alpha$ it will lead.

In Appendix we present the weak interaction Hamiltonian which is
responsible for $b \to u \bar u d$ transition and calculate the
necessary matrix elements. Using these formulas in sections 2, 3,
and 4 we study $B\to \rho\pi$, $\rho\rho$ and $\pi\pi$ decays
correspondingly and extract the values of angle $\alpha$ from the
experimental data on CP asymmetries in these decays. We conclude
in section 5 with the averaged value of $\alpha$ and a general
discussion.

\section{$\alpha$ from $\bar B_d (B_d) \to \rho^\mp \pi^\pm$}

The time dependence of the decay probabilities is given by
\cite{5}:
\begin{eqnarray}
\frac{d N(B_d(\bar B_d) \to \rho^\pm \pi^\mp)}{d t} & = & (1\pm
A_{CP}^{\rho\pi})e^{-t/\tau} [1-q(C_{\rho\pi} \pm \Delta
C_{\rho\pi}) \times \nonumber \\ & \times & \cos(\Delta m t) +
q(S_{\rho\pi}\pm \Delta S_{\rho\pi})\sin(\Delta m t)] \;\; ,
\label{2}
\end{eqnarray}
where $q=-1$ describes the case when at $t=0$ $B_d$ was produced,
while $q=1$ corresponds to $\bar B_d$ production at $t=0$. In the
case of $\Upsilon(4S) \to B_d \bar B_d$ decay the flavor of the
beauty meson which will decay to $\rho\pi$ is tagged by the charge
of a lepton in the other beauty meson semileptonic decay.  A
partner decay starts clocks as well. $\tau$ is $B_d (\bar B_d)$
life time, while $\Delta m$ is the difference of masses of $(B_d,
\bar B_d)$ system eigenstates (it equals the frequency of $B_d -
\bar B_d$ oscillations).

From Eqs. (\ref{A15}) and (\ref{A17}) we obtain:
\begin{equation}
\bar M^{-+} = A V_{ub} V_{ud}^* [1-0.07 e^{i(\delta -\alpha)}] =
AV_{ub} V_{ud}^*[1-0.07 \sin\delta + i0.07 \cos\delta] \;\; ,
\label{3}
\end{equation}
\begin{equation}
M^{-+} = B V_{ub}^* V_{ud} \;\; , \label{4}
\end{equation}
\begin{equation}
\lambda^{-+} \equiv \frac{q}{p} \frac{\bar M^{-+}}{M^{-+}} =
e^{2i\alpha} \frac{A}{B}[1-0.07 \sin\delta + i 0.07 \cos\delta]
\;\; , \label{5}
\end{equation}
where parameters $q$ and $p$ enter the expressions for $(B_d, \bar
B_d)$ eigenstates and we have substituted $\alpha = \pi/2$ in the
(small) second term in square brackets in Eq. (\ref{A15}).
Analogously we get:
\begin{equation}
\bar M^{+-} = B V_{ub} V_{ud}^* \;\; , \label{6}
\end{equation}
\begin{equation}
M^{+-} = A V_{ub}^* V_{ud}[1-0.07 e^{i(\alpha+\delta)}] = A
V_{ub}^* V_{ud} [1+0.07 \sin\delta - 0.07 i \cos\delta] \;\; ,
\label{7}
\end{equation}
\begin{equation}
\lambda^{+-} \equiv \frac{q}{p} \frac{\bar M^{+-}}{M^{+-}} =
e^{2i\alpha} \frac{B}{A} [1+0.07\sin\delta - i 0.07
\cos\delta]^{-1} \;\; . \label{8}
\end{equation}

From the expressions for the quantities $C_{\rho\pi}$ and $\Delta
C_{\rho\pi}$ \cite{5}:
\begin{equation}
C_{\rho\pi} \pm \Delta C_{\rho\pi} =
\frac{1-|\lambda^{\pm\mp}|^2}{1+|\lambda^{\pm\mp}|^2}  \label{9}
\end{equation}
we obtain:
\begin{eqnarray}
\Delta C_{\rho\pi} & = & \frac{a^2 - b^2}{a^2 + b^2} \; , \;\;
\frac{a^2}{b^2} = \frac{1+ \Delta C_{\rho\pi}}{1 - \Delta
C_{\rho\pi}} \;\; , \nonumber \\ C_{\rho\pi} & = & 0.28 \sin\delta
\frac{(a/b)^2}{(1+a^2/b^2)^2} \;\; , \label{10}
\end{eqnarray}
where $a \equiv |A|$, $b \equiv |B|$.

The Belle and BABAR averaged result for $\Delta C_{\rho\pi}$ is
\cite{6}:
\begin{equation}
\Delta C_{\rho\pi} = 0.22 \pm 0.10 \;\; , \label{11}
\end{equation}
which leads to:
\begin{equation}
\left(\frac{a}{b}\right)^2 = 1.56 \pm 0.33 \;\; . \label{12}
\end{equation}
From the averaged experimental result \cite{6}
\begin{equation}
C_{\rho\pi} = 0.31 \pm 0.10 \label{13}
\end{equation}
and Eq. (\ref{10}) we get:
\begin{equation}
\sin\delta = 4.6 \pm 1.5 \;\; . \label{14}
\end{equation}
We see that poor accuracy in the measurement of $C_{\rho\pi}$ does
not allow to get any definite information on the value of phase
$\delta$.

The next observable we wish to discuss is CP asymmetry
$A_{CP}^{\rho\pi}$:
\begin{eqnarray}
A_{CP}^{\rho\pi} & = & \frac{|M^{+-}|^2 - |\bar M^{-+}|^2 + |\bar
M^{+-}|^2 - |M^{-+}|^2}{|M^{+-}|^2 + |\bar M^{-+}|^2 + |\bar
M^{+-}|^2 + |M^{-+}|^2} = \nonumber \\ & = & 0.14 \sin\delta
\frac{(a/b)^2}{1+(a/b)^2} \;\; , \label{15}
\end{eqnarray}
which should be compared with the experimental result \cite{6}:
\begin{equation}
A_{CP}^{\rho\pi} = -0.102 \pm 0.045 \;\; . \label{16}
\end{equation}
From (\ref{15}) and (\ref{16}) we get:
\begin{equation}
\sin\delta = -1.2 \pm 0.5 \;\; , \label{17}
\end{equation}

and it differs from given in Eq.(\ref{14}) by 3.5 standard
deviations. This is the largest discrepancy we encounter in this
paper. Averaging these two numbers we obtain:
\begin{equation}
\sin\delta = -0.62 \pm 0.47 \;\; . \label{171}
\end{equation}

Finally we come to the discussion of the observables which are
sensitive to the angle $\alpha$:
\begin{equation}
S_{\rho\pi} \pm \Delta S_{\rho\pi} = \frac{2{\rm
Im}\lambda^{\pm\mp}}{1+|\lambda^{\pm\mp}|^2} \;\; , \label{18}
\end{equation}
\begin{equation}
S_{\rho\pi} = \frac{2a/b}{1+ a^2/b^2}[\sin 2\alpha
\cos\tilde\delta - 0.07\cos\delta \cos\tilde\delta - 0.07
\frac{a^2/b^2 -1}{a^2/b^2 +1} \sin\delta \sin\tilde\delta] \;\; ,
\label{19}
\end{equation}
\begin{eqnarray}
\Delta S_{\rho\pi} & = & \frac{2a/b}{1+ a^2/b^2}[-\cos 2 \alpha
\sin\tilde\delta + 0.07\cos\delta \sin\tilde\delta\sin 2\alpha -
\nonumber \\ & - & 0.07 \frac{a^2/b^2 -1}{a^2/b^2 +1} \sin\delta
\cos\tilde\delta \sin 2\alpha] \;\; , \label{20}
\end{eqnarray}
where the definition of the phase $\tilde\delta$ is $A/B \equiv
(a/b)e^{i\tilde\delta}$ and in the small terms proportional to
0.07 in the expression for $S_{\rho\pi}$ we have substituted $\cos
2 \alpha = -1$.

Let us start the analysis of the experimental data from $\Delta
S_{\rho\pi}$. According to \cite{6}:
\begin{equation}
\Delta S_{\rho\pi} = 0.09 \pm 0.13 \;\; , \label{21}
\end{equation}
which is much less than one. According to Eq. (\ref{12}) the
factor which multiplies square brackets in Eq. (\ref{19}) (and in
Eq. (\ref{20})) is very close to one, that is why it is the
expression in square brackets which should be much less than one.
The second and the third terms of this expression are really very
small and we can neglect them. What concerns the first term, it is
small when $\tilde\delta$ is close to zero or $\pi$:
\begin{equation}
\sin\tilde\delta = \Delta S_{\rho\pi} = 0.09 \pm 0.13 \;\; ,
\label{22}
\end{equation}
where a small deviation of $\cos 2 \alpha$ from -- 1 is neglected.

Now everything is ready and from Eq. (\ref{19}) we get:
\begin{equation}
\sin 2\alpha = S_{\rho\pi}/\cos\tilde\delta + 0.07 \cos\delta \;\;
, \label{23}
\end{equation}
where we omit the last term in square brackets since it is
negligibly small. In order to find angle $\alpha$ from the
experimental data \cite{6}:
\begin{equation}
S_{\rho\pi} = -0.13 \pm 0.13 \;\; , \label{24}
\end{equation}
we should determine the values of $\cos\delta$ and
$\cos\tilde\delta$.

To propagate errors from $\sin\delta$ to $\cos\delta$, we consider
gaussian distribution for $\sin\delta$ truncated to physical
region $|\sin\delta | < 1$, transform it into (non-gaussian)
distribution for
$\cos\delta$ and take an interval containing 68\% of probability.

In this way from Eq. (\ref{171}) we obtain:
\begin{equation}
|\cos\delta | = 0.88 \pm 0.12 \;\; . \label{25}
\end{equation}
The average value of
$|\cos\delta |$ appears to be close to one due to the so-called
Jacobian pick.

Concerning $\tilde\delta$ it follows from Eq. (\ref{22}) that
$|\cos\tilde\delta | = 1$ with very good accuracy. Depending on
the values of phases $\tilde\delta$ and $\delta$ we get the
following four domains for the angle $\alpha$:
\begin{eqnarray}
\tilde\delta \approx 0 \; , \;\; \delta \approx 0: & & \sin 2
\alpha = -0.13 \pm 0.13 + 0.06 \nonumber
\\
&& \alpha = 92^o \pm 4^o \label{26a}
\end{eqnarray}
\begin{eqnarray}
\tilde\delta \approx 0 \; , \;\; \delta \approx \pi: & & \sin 2
\alpha = -0.13 \pm 0.13 - 0.06 \nonumber
\\
&& \alpha = 96^o \pm 4^o \label{26b}
\end{eqnarray}
\begin{eqnarray}
\tilde\delta \approx \pi \; , \;\; \delta \approx 0: & & \sin 2
\alpha = 0.13 \pm 0.13 + 0.06 \nonumber
\\
&& \alpha = 84^o \pm 4^o \label{26c}
\end{eqnarray}
\begin{eqnarray}
\tilde\delta \approx \pi \; , \;\; \delta \approx \pi: & & \sin 2
\alpha = 0.13 \pm 0.13 - 0.06 \nonumber
\\
&& \alpha = 88^o \pm 4^o \;\; . \label{26d}
\end{eqnarray}

Thus QCD penguins split values of $\alpha$ obtained without taking
them into account: $94^o \to 92^o, 96^o$; $86^o \to 84^o, 88^o$.
If BSS mechanism is valid, then only the domains given by
Eqs.(\ref{26a}) and (\ref{26c}) remain (see also \cite{77}).

\section{$\alpha$ from $\bar B_d (B_d) \to \rho^+ \rho^-$}

The time dependence of CP violating asymmetry is described by the
following formula:
\begin{equation}
a_{CP}(t) \equiv  \frac{\frac{dN(\bar B_d \to \rho_L^+
\rho_L^-)}{dt} - \frac{dN(B_d \to \rho_L^+
\rho_L^-)}{dt}}{\frac{dN(\bar B_d \to \rho_L^+ \rho_L^-)}{dt} +
\frac{dN(B_d \to \rho_L^+ \rho_L^-)}{dt}} = -C_{\rho\rho}
\cos(\Delta m t) + S_{\rho\rho}\sin(\Delta m t) \;\; . \label{27}
\end{equation}

Let us remind that the longitudinal polarization fraction $f_L =
0.98 \pm 0.01 \pm 0.02$ \cite{7};
$f_L = 0.95 \pm 0.03 \pm 0.03$ \cite{78}
 and its closeness to one greatly
simplify the extraction of CPV parameters from $B_d (\bar B_d) \to
\rho^+ \rho^-$ decay data. These parameters are given by the
following expressions:
\begin{equation}
C_{\rho\rho} =
\frac{1-|\lambda_{\rho\rho}|^2}{1+|\lambda_{\rho\rho}|^2} \; ,
\;\; S_{\rho\rho} = \frac{2 {\rm Im}
\lambda_{\rho\rho}}{1+|\lambda_{\rho\rho}|^2} \;\; . \label{28}
\end{equation}
From Eq. (\ref{A14}) we obtain:
\begin{equation}
\lambda_{\rho\rho} \equiv \frac{q}{p} \frac{\bar
M_{\rho\rho}}{M_{\rho\rho}} = e^{2i\alpha} \frac{1-0.07
e^{i(\delta - \alpha)}}{1-0.07 e^{i(\delta + \alpha)}} =
e^{2i\alpha}(1+0.14 i \sin\alpha e^{i\delta}) \;\; , \label{29}
\end{equation}
where we use the same letter $\delta$ for FSI phase difference of
the amplitudes generated by penguin and tree diagrams as in the
case of $B\to \rho\pi$ decays. These differences would be really
the same if BSS mechanism dominates.

Comparing the averaged experimental result for $C_{\rho\rho}$
\begin{equation}
C_{\rho\rho} = -0.03 \pm 0.18 \pm 0.09 \;\;\;\cite{7},   \label{30}
\end{equation}

\begin{equation}
C_{\rho\rho} = 0.0 \pm 0.30 \pm 0.10 \;\;\;\cite{78} ;  \label{300}
\end{equation}
\begin{equation}
C_{\rho\rho}^{exp} = -0.02 \pm 0.17   \label{301}
\end{equation}
with the theoretical expression which follows from Eqs.
(\ref{28}), (\ref{29})
\begin{equation}
C_{\rho\rho} = 0.14 \sin\alpha \sin\delta \approx 0.14 \sin\delta
\label{31}
\end{equation}
we get:
\begin{equation}
\sin\delta = -0.15 \pm 1.2  \; ; |\cos\delta | = 0.88 \pm
0.12 \;\; , \label{32}
\end{equation}
where the same procedure of error propagation as in the case of
$B\to \rho\pi$ was used and we sum statistical and systematic
errors of $\sin\delta$ as independent.

From Eq. (\ref{28}) we obtain:
\begin{equation}
S_{\rho\rho} = \sin 2 \alpha + 0.14 \sin\alpha \cos 2\alpha
\cos\delta = \sin 2\alpha - 0.14 \cos\delta \;\; . \label{33}
\end{equation}
According to the recent measurements:
\begin{equation}
S_{\rho\rho} = -0.33 \pm 0.26 \;\; \cite{7}, \label{34}
\end{equation}

\begin{equation}
S_{\rho\rho} = 0.09 \pm 0.43 \;\; \cite{78}, \label{340}
\end{equation}

\begin{equation}
S_{\rho\rho}^{exp} = -0.21 \pm 0.22 \;\; , \label{341}
\end{equation}

and we get two domains for $\alpha$:
\begin{equation}
\delta \approx 0: ~ \alpha = 92^o \pm 7^o \label{35a}
\end{equation}
\begin{equation}
\delta \approx \pi: ~ \alpha = 100^o \pm 7^o \;\; . \label{35b}
\end{equation}
Just as in the case of $B\to \rho\pi$ decays only the first domain
remains if $|\delta | < \pi/2$ \cite{77}.

Let us note that using the isospin analysis (which allows to prove
the smallness of the penguin contribution) it
was obtained:
\begin{equation}
\alpha = 100^o \pm 13^o \;\; \cite{7}, \label{36}
\end{equation}
\begin{equation}
\alpha = 87^o \pm 17^o \;\; \cite{78}, \label{360}
\end{equation}
where the error is mainly due to the uncertainty of the penguin
contribution. Extracting this uncertainty and averaging last
two numbers we get:
\begin{equation}
\alpha = 96^o \pm 7^o (exp) \pm 11^o (penguin) \;\; . \label{361}
\end{equation}
\section{$\alpha$ from $\bar B_d(B_d) \to \pi^+ \pi^-$}

The time dependence of CP violating asymmetry is given by the
formula analogous to Eq. (\ref{27}) with the evident substitution
of $\pi$ instead of $\rho$. Eq. (\ref{28}) with the same
substitution is valid as well, while for the quantity
$\lambda_{\pi\pi}$ from Eq. (\ref{A13}) we obtain:
\begin{equation}
\lambda_{\pi\pi} \equiv \frac{q}{p} \frac{\bar M_{\pi^+
\pi^-}}{M_{\pi^+ \pi^-}} = e^{2i\alpha}(1+ 0.28 i \sin\alpha
e^{i\delta}) \;\; , \label{37}
\end{equation}
and what concerns letter $\delta$ we should repeat the comment
made after Eq. (\ref{29}).

The experimental data of BABAR and Belle for CPV parameters
$S_{\pi\pi}$ and $C_{\pi\pi}$  were controversial  though at
present (with the latest Belle results) the divergence diminishes.
In view of this we will perform a two step analysis, taking at the
beginning only BABAR results and then the averaged results of two
collaborations.

Comparing the theoretical expression
\begin{equation}
C_{\pi\pi} = 0.28 \sin\alpha\sin\delta \approx 0.28 \sin\delta
\label{38}
\end{equation}
with BABAR result \cite{6}
\begin{equation}
C_{\pi\pi}^{\rm BABAR} = -0.09 \pm 0.15 \label{39}
\end{equation}
we get:
\begin{equation}
\sin\delta = -0.32 \pm 0.54 \;\; , \label{40}
\end{equation}
while for $S_{\pi\pi}$ we have:
\begin{equation}
S_{\pi\pi} = \sin 2 \alpha - 0.28 \cos\delta \; , \;\;
S_{\pi\pi}^{\rm BABAR} = -0.30 \pm 0.17 \;\; . \label{41}
\end{equation}

From Eq. (\ref{40}) we get:
\begin{equation}
|\cos\delta | = 0.9 \pm 0.1 \label{411}
\end{equation}
and two domains of $\alpha$ corresponding to two signs of
$\cos\delta$\footnote{Assuming $|\delta | < \pi/2$ \cite{77} we
would get only the first domain.}:
\begin{equation}
\delta \approx 0: ~ \alpha = 91^o \pm 5^o ({\rm exp}) \pm 1^o
({\rm theor}) \label{42a}
\end{equation}
\begin{equation}
\delta \approx \pi: ~ \alpha = 107^o \pm 5^o ({\rm exp}) \pm 1^o
({\rm theor}) \;\; . \label{42b}
\end{equation}

Belle result:
\begin{equation}
C_{\pi\pi}^{\rm Belle} = -0.56 \pm 0.13 \label{43}
\end{equation}
deviates by $2.5\sigma$ from BABAR and (if correct) would require
considerably larger $P/T$ ratio than we use in our paper.

Finally, averaging (\ref{39}) and (\ref{43}) one gets \cite{6}:
\begin{equation}
C_{\pi\pi}^{\rm ex} = -0.37 \pm 0.10 \;\; , \label{44}
\end{equation}
and comparing with the theoretical  expression (\ref{38}) we
obtain:
\begin{equation}
\sin\delta = -1.32 \pm 0.35 \;\; , \label{45}
\end{equation}
which leads to:
\begin{equation}
|\cos\delta | = 0.55 \pm 0.25 \;\; . \label{455}
\end{equation}

From the averaged experimental result \cite{6}:
\begin{equation}
S_{\pi\pi}^{\rm ex} = -0.50 \pm 0.12 ({\rm exp}) \label{46}
\end{equation}
we get two domains\footnote{Assuming $|\delta | < \pi/2$ \cite{77}
which follows from BSS mechanism of $\delta$ generation, we would
get the first domain only.}:
\begin{equation}
\delta \approx 0: ~ \alpha = 100^o \pm 4^o ({\rm exp}) \pm 2^o
({\rm theor}) \label{47a}
\end{equation}
\begin{equation}
\delta \approx \pi: ~ \alpha = 110^o \pm 4^o ({\rm exp}) \pm 2^o
({\rm theor}) \;\; . \label{47b}
\end{equation}

\section{Conclusions}

We have analyzed CPV asymmetries in $B_d(\bar B_d) \to \rho^\pm
\pi^\mp$, $\rho_L^+ \rho_L^-$ and $\pi^+ \pi^-$ decays induced by
the charmless strangeless $b$-quark decay $b\to u \bar u d$. This
decay can proceed through a tree or penguin diagram. As it was
noted in \cite{1} when the penguin diagram is neglected, one
obtains the values of the unitarity triangle angle $\alpha$ from
CPV asymmetries in these decays which are consistent with each
other as well as with the value of $\alpha$ which follows from the
global CKM fit. However, in order to determine the theoretical
accuracy of $\alpha$ extracted from the decays under study one
should take the penguin amplitude into account. This was done in
the present paper, where the moduli of penguin over tree ratios
were calculated with the help of the factorization:
\begin{equation}
<M_1 M_2 |j_1 j_2|B> = <M_1 |j_1|B> <M_2 |j_2|0> \;\; , \label{48}
\end{equation}
while FSI phase shifts between the tree and penguin amplitudes
were extracted from experimental data.

In order to determine numerical value of $\alpha$ one should
average the values which follow from the considered decays. Since
the phase shifts $\delta$ can be different in $B\to \rho\rho,
\pi\pi$ and $\rho\pi$ decays, we get too many possibilities. That
is why let us limit ourselves to the theoretically motivated
case $|\delta | < \pi/2$.

Averaging Eqs. (\ref{35a}) and (\ref{47a}) we obtain:
\begin{equation}
\delta \approx 0: ~ \alpha_{\rho\rho, \pi\pi} = 98^o \pm 4^o \;\;
. \label{58}
\end{equation}
Averaging it with (\ref{26a}) and (\ref{26c}) we get two
possibilities:
\begin{eqnarray}
\alpha_{b\to u\bar u d} & = & 95^o \pm 3^o \; ,  \;\; {\rm or}
\nonumber \\ \alpha_{b\to u\bar u d} & = & 91^o \pm 3^o \; ,
\label{59}
\end{eqnarray}
and the last one corresponds to the smallest possible value of $\alpha$. In the case $|\delta_i | > \pi/2$, $\tilde\delta = 0$ averaging Eqs.(\ref{26b}), (\ref{35b}), (\ref{47b}) we get the largest possible value: $\alpha_{b\to u\bar u d} = 102^o \pm 3^o$.

The global fit results for $\alpha$ are:
\begin{equation}
\alpha_{\rm UTfit}^{\rm\cite{8}} = 94^o \pm 8^o \; , \;\;
\alpha_{\rm CKMfitter}^{\rm \cite{9}} = 94 \pm 10^o \;\; .
\label{55}
\end{equation}

Thus the accuracy of the present day knowledge of $\alpha$ can be
close to that of $\beta$:
\begin{equation}
\beta = 23^o \pm 2^o \;\; . \label{56}
\end{equation}

I am grateful to  A.V.~Fedotov  for great help in the treatment of
the experimental data. I would like to thank A.E.~Bondar and
M.B.~Voloshin for useful remarks. This work was partially
supported by grants NSh-2328.2003.2 and RFBR 05-02-17203.

\newpage

\begin{center}

{\bf Appendix}

\end{center}

\setcounter{equation}{0} \def\theequation{A\arabic{equation}}

\bigskip

The strong interaction renormalization of the tree Hamiltonian
which describes the beauty hadrons weak decays is much smaller
than for the case of the strange particle decays since the masses
of beauty hadrons are much closer to $M_W$ in the logarithmic
scale. In the leading logarithmic approximation for operators
$O_1$ and $O_2$ we have:
\begin{eqnarray}
\hat H_{1,2} &=& \frac{G_F}{\sqrt 2} V_{ub} V_{ud}^*
\left\{\left[\frac{\alpha_S(m_b)}{\alpha_S(M_W)}\right]^{4/b}
\left[\bar u \gamma_\alpha(1+\gamma_5)b \bar d
\gamma_\alpha/1+\gamma_5)u - \right.\right. \nonumber \\ &-&
\left. \bar d \gamma_\alpha(1+\gamma_5) b \bar u
\gamma_\alpha(1+\gamma_5)u\right] +
\left[\frac{\alpha_S(m_b)}{\alpha_S(M_W)}\right]^{-2/b} [\bar u
\gamma_\alpha(1+\gamma_5) b \times \nonumber \\ & \times & \left.
\bar d \gamma_\alpha(1+\gamma_5) u + \bar d
\gamma_\alpha(1+\gamma_5)b \bar u \gamma_\alpha(1+\gamma_5)u]
\right\} \label{A1}
\end{eqnarray}
and substituting $\alpha_S(M_W) = 0.12$, $\alpha_S(m_b) = 0.2$, $b
= 23/3$ we get:
\begin{eqnarray}
\hat H_{1,2} & = & \frac{G_F}{\sqrt 2} V_{ub} V_{ud}^* \{1.1 \bar
u \gamma_\alpha(1+\gamma_5)b \bar d \gamma_\alpha(1+\gamma_5)u -
\nonumber \\ & - & 0.2 \bar d \gamma_\alpha(1+\gamma_5)b \bar u
\gamma_\alpha(1+\gamma_5)u \} \;\; . \label{A2}
\end{eqnarray}

NLO calculations confirm and refine this result. From Table 1 of
\cite{10} for the value $\Lambda_4 = 280$ MeV (which corresponds
to $\alpha_S(M_Z) = 0.118$) we get 1.14 instead of our 1.1 and
-0.31 instead of our -0.2.

At one loop the following QCD penguin operator is generated:
\begin{eqnarray}
\hat H_{3-6} & = & -\frac{G_F}{\sqrt 2}(V_{cb}V_{cd}^* + V_{ub}
V_{ud}^*) \frac{\alpha_S(m_b)}{12\pi}
\ln\left(\frac{M_W}{m_b}\right)^2 (\bar d
\gamma_\mu(1+\gamma_5)\vec\lambda b) \times \nonumber \\ & \times
& (\bar u \gamma_\mu \vec\lambda u + \bar d \gamma_\mu \vec
\lambda d) = + \frac{G_F}{\sqrt 2}V_{tb} V_{td}^* 0.03 \{
-\frac{2}{3} \bar d \gamma_\alpha(1+\gamma_5)b \times \nonumber \\
& \times & (\bar u \gamma_\alpha u + \bar d \gamma_\alpha d)
+2(\bar d_a \gamma_\alpha(1+\gamma_5)b^c)(\bar u_c \gamma_\alpha
u^\alpha + \bar d_c \gamma_\alpha d^a)\} \;\; , \label{A3}
\end{eqnarray}
where $\vec\lambda$ are eight colour SU(3) Gell-Mann matrices, and
Fierz identity $\vec\lambda_{ab} \vec\lambda_{cd} = -2/3
\delta_{ab} \delta_{cd} + 2\delta_{ad} \delta_{bc}$ as well as
unitarity relation $V_{cb}V_{cd}^* + V_{ub}V_{ud}^* = -V_{tb}
V_{td}^*$ were used.

Substituting $\bar q\gamma_\alpha q = \frac{1}{2}\bar
q\gamma_\alpha(1+\gamma_5)q + \frac{1}{2}\bar
q\gamma_\alpha(1-\gamma_5)q$ we find the renormalization factors
+0.01 and -0.03 for operators $O_3$, $O_5$ and $O_4$, $O_6$
respectively. At NLO the renormalization factors for the operators
in which only the left-handed  quarks are involved ($O_3, O_4$)
are different from those for the operators in which both left- and
right-handed quarks participate ($O_5, O_6$). From the same Table
1 of \cite{10} we get 0.016 and 0.010 instead of 0.01 and -0.036
and -0.045 instead of -0.03.

Finally, the effective Hamiltonian which describes the charmless
strangeless $\bar B_d$ decays looks like:
\begin{equation}
\hat H = \frac{G_F}{\sqrt 2} \left[V_{ub} V_{ud}^* (c_1 O_1 +c_2
O_2) - V_{tb}V_{td}^*(c_3 O_3 + c_4 O_4 + c_5 O_5 +c_6 O_6)
\right] \;\; , \label{A4}
\end{equation}
\begin{equation}
\begin{array}{ll}
O_1 = \bar u \gamma_\alpha(1+\gamma_5)b \bar
d\gamma_\alpha(1+\gamma_5)u & c_1 = 1.14 \; ,  \\ O_2 = \bar d
\gamma_\alpha(1+\gamma_5)b \bar u \gamma_\alpha(1+\gamma_5)u & c_2
= -0.31 \; ,  \\ O_3 = \bar d \gamma_\alpha(1+\gamma_5)b [\bar u
\gamma_\alpha(1+\gamma_5) u + \bar d \gamma_\alpha(1+ \gamma_5)d]
& c_3 = 0.016 \; ,
\\ O_4 = \bar d_a \gamma_\alpha(1+\gamma_5)b^c [\bar u_c
\gamma_\alpha(1+\gamma_5)u^a + \bar d_c
\gamma_\alpha(1+\gamma_5)d^a] & c_4 = -0.036 \; ,  \\ O_5 = \bar d
\gamma_\alpha(1+\gamma_5)b[\bar u \gamma_\alpha(1-\gamma_5)u +
\bar d \gamma_\alpha(1-\gamma_5)d] & c_5 = 0.010 \; ,  \\ O_6 =
\bar d_a \gamma_\alpha(1+\gamma_5)b^c [\bar u_c
\gamma_\alpha(1-\gamma_5) u^a + \bar d_c
\gamma_\alpha(1-\gamma_5)d^a] & c_6 = -0.045 \; , \end{array}
\label{A5}
\end{equation}
and the complex conjugate Hamiltonian describes $B_d$ decays.

Our next task is to calculate the matrix elements of $\hat H$
between $\bar B_d$ and $\rho^\pm \pi^\mp$, $\rho^+ \rho^-$ and
$\pi^+ \pi^-$ states, which is the most difficult part of the job.
We will present the matrix elements of 4-fermion operators as the
product of matrix elements of two 2-fermion operators between
$\bar B_d$ and a light meson and vacuum and another light meson.
The validity of this factorization is questionable; in particular,
in this approach the FSI phases due to the light meson
rescattering vanish identically. However, we found the statement
in the literature that the corrections to the factorization
formulas are small, being proportional to $\Lambda/m_b$ or powers
of $\alpha_S(m_b)$ \cite{11}\footnote{Since $q^2 \equiv (P_{B_d} -
P_{\pi,\rho})^2 = O(m_\pi^2, m_\rho^2)$ one can argue that the
large distance contributions invalidate the factorization
formula.}. In any case nowadays factorization is the only way to
get expressions for the decay amplitudes from the fundamental
Hamiltonian.

Since we are interested in $\bar B_d$ decays to charged mesons and
we will factorize 4-fermion operators, let us present Eqs.
(\ref{A4}), (\ref{A5}) in the following form \cite{2}:
\begin{eqnarray}
\hat H &=& \frac{G_F}{\sqrt 2} V_{ub} V_{ud}^* \{ a_1 \bar u
\gamma_\alpha(1+\gamma_5) b \bar d \gamma_\alpha(1+\gamma_5)u -
\nonumber \\ &-& \frac{V_{tb}V_{td}^*}{V_{ub}V_{ud}^*} [a_4 \bar u
\gamma_\alpha(1+\gamma_5)b \bar d \gamma_\alpha(1+\gamma_5)u - \\
&-& 2a_6 \bar u(1+\gamma_5)b \bar d(1-\gamma_5)u]\} \;\; ,
\nonumber \label{A6}
\end{eqnarray}
where $a_1 =c_1 + \frac{1}{3}c_2 = 1.04$, $a_4 = c_4 +
\frac{1}{3}c_3 = -0.031$, $a_6 = c_6 + \frac{1}{3} c_5 = -0.042$
and Fierz identities $\bar\psi \gamma_\alpha(1+\gamma_5)\varphi
\bar\chi \gamma_\alpha(1+\gamma_5)\eta = \bar\psi
\gamma_\alpha(1+\gamma_5)\eta \bar\chi
\gamma_\alpha(1+\gamma_5)\varphi$, $\bar\psi
\gamma_\alpha(1+\gamma_5)\varphi \bar\chi
\gamma_\alpha(1-\gamma_5)\eta = -2\bar\psi(1-\gamma_5)\eta
\bar\chi (1+\gamma_5)\varphi$ were used.

The matrix elements we are interested in were calculated in paper
\cite{2} assuming factorization. Up to a common factor which
includes constant $f_\pi$ and $B\to\pi$ transition formfactor
$f_0(m_\pi^2)$ for the amplitude of $\bar B_d \to \pi^+ \pi^-$
decay it was obtained:
\begin{equation}
\frac{M(\bar B_d \to \pi^+ \pi^-)}{V_{ub}V_{ud}^*} \sim a_1 -
\frac{V_{tb} V_{td}^*}{V_{ub} V_{ud}^*} \left[ a_4 + \frac{2m_\pi
^2}{(m_u + m_d)(m_b - m_u)} a_6 \right] e^{i\delta} \;\; ,
\label{A7}
\end{equation}
where we use the result of \cite{2} and take into account the
difference of the rescattering phases of the tree ($\sim  a_1$)
and the penguin ($\sim a_4$ and $\sim a_6$) amplitudes $\delta$.
One evident source of this phase is the imaginary part of the
penguin diagrams with intermediate $u$- and $c$-quarks. Let us
demonstrate that only the last one should be taken into account in
(\ref{A7}):
\begin{eqnarray}
M(\bar B_d & \to & \pi^+ \pi^-)  \sim  V_{ub} V_{ud}^*(T+ P(m_u))
+ V_{cb} V_{cd}^* P(m_c) + V_{tb} V_{td}^* P(m_t) = \nonumber \\ &
= & V_{ub} V_{ud}^* [T + P(m_u) - P(m_c)] -V_{tb} V_{td}^* [P(m_c)
- P(m_t)] \;\; . \label{A8}
\end{eqnarray}
As we are interested in CP asymmetries we should calculate
$\lambda_{\pi^+ \pi^-} = e^{-2i\beta} M(\bar B_d \to \pi^+
\pi^-)/M(B_d \to \pi^+ \pi^-)$:
\begin{eqnarray}
\lambda & = & e^{-2 i\beta -2i\gamma} \frac{1+ \frac{P(m_u) -
P(m_c)}{T} - \frac{V_{tb} V_{td}^*}{V_{ub} V_{ud}^*} \frac{P(m_c)
- P(m_t)}{T}}{1+ \frac{P(m_u) - P(m_c)}{T} - \frac{V_{tb}^*
V_{td}}{V_{ub}^* V_{ud}} \frac{P(m_c) - P(m_t)}{T}} = \nonumber \\
& = & e^{2i\alpha}\left[ 1+(e^{i(\alpha -\pi)} -e^{i(\pi
-\alpha)}) \frac{\sin\gamma}{\sin\beta} \frac{P(m_c) - P(m_t)}{T}
\right] \;\; , \label{A9}
\end{eqnarray}
where $\alpha$, $\beta$ and $\gamma$ are the angles of the
unitarity triangle.

Thus the absorptive part of $P(m_c)$ contributes to $\delta$.

Since the penguin operator $P(m_c)$ equals the correlator of two
vector currents, one immediately picks up its imaginary part from
the textbooks on QED. It depends on the gluon momentum transfer
and when the square of this momentum transfer is much larger than
$4m_c^2$, we have:
\begin{equation}
\frac{P(m_c)}{T} \sim -\ln \frac{M_W^2}{m_b^2} - i\pi \equiv -
\left|\frac{P}{T}\right| e^{i\delta} \; , \;\; \delta \approx 30^o
\;\; . \label{A10}
\end{equation}
Since $u$- and $\bar u$-quarks to which gluon decays go to
different light mesons, the value of the momentum transfer squared
is determined by these mesons wave functions. It varies between
$m_b^2$ and zero. Thus we see that the mechanism suggested in
\cite{3} leads to the small positive value of $\delta$:
\begin{equation}
\delta \la 30^o \;\; . \label{A11}
\end{equation}

If BSS mechanism determines the value of $\delta$, it would
confirm the validity of our approach. If, on the contrary, the
large distance rescattering of light hadrons changes $\delta$
substantially, one should await large corrections to the
dispersive part (the ratio ($P/T$)) as well. In the present paper
we will allow $\delta$ to vary between zero and $2\pi$, but we use
the expressions analogous to (\ref{A7}) for the decay amplitudes.

Let us return to Eq. (\ref{A7}). With the help of the following
equation:
\begin{equation}
\frac{V_{td}^* V_{tb}}{V_{ud}^* V_{ub}} = e^{i(\pi -\alpha)}
\frac{\sin\gamma}{\sin\beta} \label{A12}
\end{equation}
and using the numerical values $m_u + m_d = 11$ MeV, $m_b = 4.5$
GeV we obtain:
\begin{eqnarray}
M(\bar B_d \to \pi^+ \pi^-) & \sim & V_{ub} V_{ud}^* \left[
1-e^{i(\pi-\alpha)} \frac{\sin\gamma}{\sin\beta} (-0.06)
e^{i\delta} \right] = \nonumber \\ & = & V_{ub} V_{ud}^* [1+ 0.14
e^{i(\pi -\alpha +\delta)}] \;\; , \label{A13}
\end{eqnarray}
where $\beta = 23^o$ and $\gamma = 63^o$ were substituted (we are
using the value of $\gamma$ from the global CKM fit in order to
estimate a small correction to the amplitude) and we put $a_1$
equal to one. In Section 4 we
analyze the experimental data on CP asymmetries in $\bar B_d (B_d)
\to \pi^+ \pi^-$ decays.

Coming to $\bar B_d(B_d) \to \rho^+ \rho^-$ decays we should
calculate the corresponding matrix element of the Hamiltonian
presented in (A6). Factorizing 4-quark operators we observe that
the term proportional to $a_6$ vanishes, since \\ $<\rho |\bar
d(1-\gamma_5) u|0> = 0$: the (pseudo) scalar current cannot
produce a vector meson from vacuum. That is why instead of Eq.
(\ref{A7}) we get (see also \cite{2}):
\begin{eqnarray}
\frac{M(\bar B_d \to \rho_L^+ \rho_L^-)}{V_{ub} V_{ud}^*} & \sim &
a_1 - \frac{V_{tb}V_{td}^*}{V_{ub}V_{ud}^*} a_4 e^{i\delta} \;\; ,
\nonumber \\ M(\bar B_d \to \rho_L^+ \rho_L^-) & \sim & V_{ub}
V_{ud}^* [1+ 0.07 e^{i(\pi -\alpha +\delta)}] \;\; . \label{A14}
\end{eqnarray}

The production of transversely polarized $\rho$-mesons by the
vector current is suppressed as $(m_\rho /m_B)^2$, and the
experimental data confirm the dominance of $\rho_L$.

If BSS mechanism is responsible for the phase $\delta$, then it
should be the same as in Eq. (\ref{A13}).  Eq. (\ref{A14}) is used
in Section 3 to extract the value of angle $\alpha$.

Our last problem is the calculation of the amplitudes $M(\bar B_d
\to \rho^\mp \pi^\pm) \equiv \bar M^{\mp\pm}$. The amplitude $\bar
M^{-+}$ corresponds to $\rho^-$ production from vacuum by $(\bar d
u)$ current, so the term proportional to $a_6$ does not contribute
to it:
\begin{eqnarray}
\frac{\bar M^{-+}}{V_{ub}V_{ud}^*} & = & A\left[ a_1 -
\frac{V_{tb} V_{td}^*}{V_{ub} V_{ud}^*} a_4 e^{i\delta_-} \right]
\;\; ; \nonumber \\ \bar M^{-+} & = & A V_{ub} V_{ud}^* \left[ 1+
0.07 e^{i(\pi -\alpha +\delta_-)}\right] \;\; , \label{A15}
\end{eqnarray}
where $A$ is the complex number.

In the case of the amplitude $\bar M^{+-}$ it is $\pi^-$ which is
produced from vacuum by $(\bar d u)$ current, so the term
proportional to $a_6$ contributes as well:
\begin{equation}
\frac{\bar M^{+-}}{V_{ub} V_{ud}^*} = B\left\{ a_1 - \frac{V_{tb}
V_{td}^*}{V_{ub} V_{ud}^*} \left[ a_4 - \frac{2m_\pi ^2}{(m_b +
m_u)(m_u + m_d)} a_6\right] e^{i\delta_+}\right\} \;\; ,
\label{A16}
\end{equation}
see \cite{2}. Here $B$ is the complex number. Unlike the case of
$\bar B_d \to \pi^+ \pi^-$ decay the terms proportional to $a_4$
and $a_6$ have opposite signs and as a result the expression in
square brackets with good accuracy equals zero, leading to:
\begin{equation}
\bar M^{+-} = B V_{ub} V_{ud}^* \;\; , \label{A17}
\end{equation}
and the penguin pollution is absent ($a_1$ is omitted since it is
very close to one).

The amplitudes of $B_d$ meson decays, $M^{-+}$ and $M^{+-}$, equal
to $\bar M^{+-}$ and $\bar M^{-+}$ correspondingly with the
complex conjugate CKM matrix elements. We will use formulas
(\ref{A15}) and (\ref{A17}) in order to determine angle $\alpha$
in Section 2 and will omit index ``--'' from $\delta_-$, since
$\delta_+$ did not enter Eq. (\ref{A17}).

Thus the penguin pollution is minimal in $\rho\pi$ mode,
intermediate in $\rho\rho$ mode and maximal in $\pi\pi$ mode (as
it was noted in \cite{2}).

\end{document}